\documentclass{ws-procs9x6}

\newcommand{\PO}{\rm l \! P }

\newcommand{\xpom}{x_{\PO} }

\begin{document}

\title{H1 Diffractive Structure Functions Measurement from new data}

\author{E.~SAUVAN \footnote{\uppercase{O}n behalf of the \uppercase{H}1 \uppercase{C}ollaboration}}

\address{Centre de Physique des Particules de Marseille \\ 
163 Avenue de Luminy F-13288 Marseille cedex 9, France\\ 
E-mail: sauvan@cppm.in2p3.fr}

\maketitle

\abstracts{
New measurements of the reduced cross section $\sigma_r^{D(3)}$ for the diffractive process $ep \rightarrow eXY$ in the kinematic domain  $12 \leq Q^2 \leq 90$~GeV$^2$,
$0.01 \leq \beta \leq 0.65$ and $\xpom<0.1$ are presented.
Data events recorded by the H1 detector during the years 1999--2000 and 2004 have been used, corresponding to a total integrated luminosity of 68~pb$^{-1}$.
The measurements are derived in the same range as previous H1 data, namely $M_Y < 1.6$ GeV and $|t| < 1.0$ GeV$^2$. 
Two different analysis methods, rapidity gap and $M_X$, are used and similar results are obtained in the kinematic domain of overlap.
Finally, together with previous data, the diffractive structure function measurements are analysed with a model based on the dipole formulation of diffractive scattering. It is found to give a very good description of the data over the whole kinematic range.}
%

\section{Introduction}

At low $x$ in deep inelastic scaterring (DIS) at HERA, approximately 10~\% of the events are of the type $ep \rightarrow eXp$, where the final state proton carries more than 95~\% of the proton beam energy.
For these processes, a photon virtuality $Q^2$, coupled to the incoming lepton, undergoes a strong interaction with the proton to form an hadronic final state system $X$ of mass $M_X$ separated by a large rapidity gap (LRG) from the leading proton. In such a reaction no net quantum number are exchanged and a fraction $\xpom$ of the proton longitudinal momentum is transferred to the system $X$. In addition the virtual photon couples to a quark carrying a fraction $\beta = \frac{x}{\xpom}$ of the exchanged momentum.

\section{Large rapidity gap measurements}

\begin{figure}[ht]
\centerline{\epsfxsize=3.2in\epsfbox{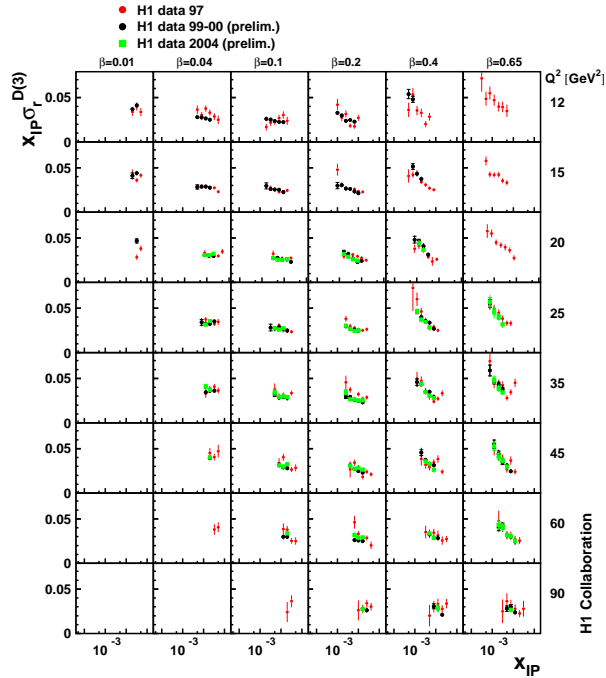}}   
\caption{Reduced cross sections $\xpom \sigma_{red}^{D(3)}$ of this
analysis for the years 1999--2000 and 2004, compared to published H1 measurements using data sample from 1997. \label{fig:f2d3_9904}}
\end{figure}

Two sets of data taken during the years 1999--2000 and 2004 and corresponding each to an integrated luminosity of 34 pb$^{-1}$ are used.
Diffractive events are selected by requiring the presence of a LRG with $\eta_{max} < 3.2$ and no activity in forward H1 detectors.
The reduced cross section $\sigma_r^{D(3)}$ measured using both samples are presented in Figure~\ref{fig:f2d3_9904} and compared to previous published H1 results \cite{h1_97paper}. 
A good agreement between the three data sets is observed, confirming with a larger statistic previous H1 measurements.

\section{The $M_X$ method}

\begin{figure}[ht]
\centerline{\epsfxsize=7cm\epsfbox{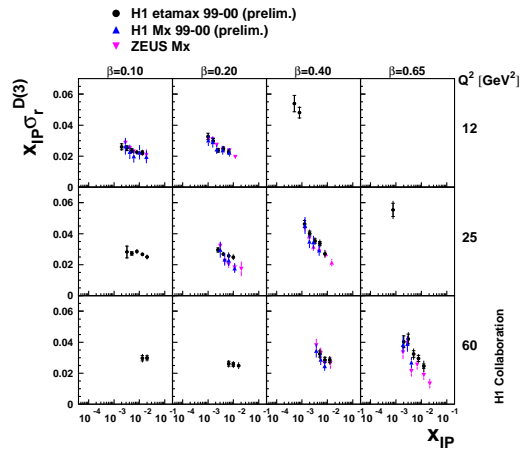}}   
\caption{ Comparisons of the diffractive reduced cross sections obtained
using the LRG and $M_X$ methods with 1999--2000 data and with measurements from the ZEUS Collaboration. \label{fig:Mx_A}}
\end{figure}

\begin{figure}[ht]
\centerline{\epsfxsize=5.5cm\epsfbox{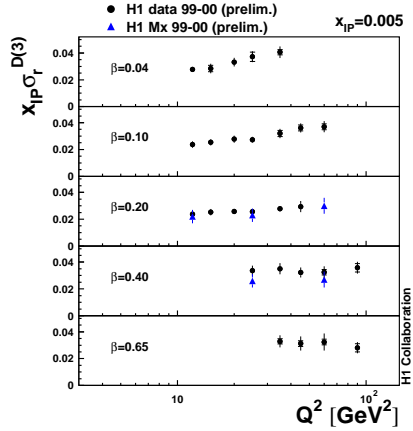}}   
\caption{Scaling violations of the diffractive cross section measurements
for both LRG and $M_X$ methods (see text). \label{fig:Mx_B}}
\end{figure}

The diffractive contribution can be extracted using another method based on the specific shape of the $\ln M_X^2$ spectrum for diffractive events, as proposed in \cite{ZEUS}. This method is applied here for the first time to H1 data. 
The reduced cross section measured using the LRG and $M_X$ method and the same 1999--2000 data set are compared in Figure~\ref{fig:Mx_A}. We notice the good agreement between
both approaches in the kinematic range accessed by this analysis.
This is also the case when we convert the ZEUS measurements in the same form,
as illustrated on Figure.~\ref{fig:Mx_A}.
Scaling violations for both methods are presented on Fig.~\ref{fig:Mx_B} and are found to agree.

\section{Comparison with the BEKW model}

\begin{figure}[ht]
\centerline{\epsfxsize=6cm\epsfbox{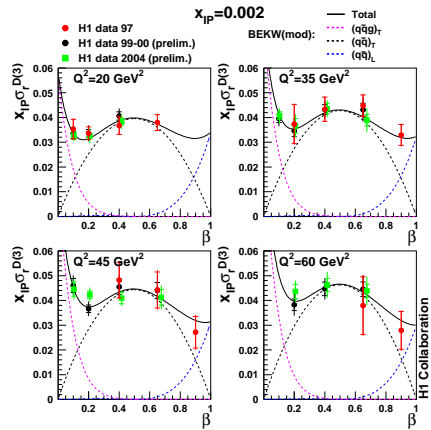}}   
\caption{ Comparisons of the diffractive cross section values
of this analysis (1999--2004) and of 1997 from \protect\cite{h1_97paper}
with the prediction of the two-gluon exchange model. Results are presented as a function of $\beta$, for 
different $Q^2$ values, for a fixed $\xpom=0.002$ value.\label{fig:BEKW}}
\end{figure}

The BEKW model \cite{bartels} provides a general parametrisation of the diffractive structure function in terms of three main contributions. 
The first contribution $F_2^{D(3)}(q \bar{q}_{_T})$ corresponds to the
diffractive scattering of the
transverse $q \bar{q}$ component of the photon. The second,
$F_2^{D(3)}(q \bar{q} g_{_T})$, corresponds to
the transverse $q \bar{q} g$
component, and the third, $F_2^{D(3)}(q \bar{q}_{_L})$, which gives a
higher twist contribution, corresponds to the longitudinal $q \bar{q}$
component.
A fit has been performed to this structure function including present $\sigma_r^{D(3)}$ measurements as well as published diffractive cross-sections from \cite{h1_97paper} which extends over a larger $Q^2$ and $\beta$ range.
A good description of all data sets together is observed. 
As an illustration, the contributions from longitudinal and transverse $q \bar{q}$ terms as well as $q \bar{q} g$ term resulting from the fit are presented on Figure \ref{fig:BEKW} and compared to data points.

\end{document}